\documentclass[pdflatex,sn-mathphys-num]{sn-jnl}
\usepackage{graphicx}%
\usepackage{multirow}%
\usepackage{amsmath,amssymb,amsfonts}%
\usepackage{amsthm}%
\usepackage{mathrsfs}%
\usepackage[title]{appendix}%
\usepackage{xcolor}%
\usepackage{textcomp}%
\usepackage{manyfoot}%
\usepackage{booktabs}%
\usepackage{algorithm}%
\usepackage{algorithmicx}%
\usepackage{algpseudocode}%
\usepackage{listings}%
\usepackage{csquotes}

\theoremstyle{thmstyleone}%

\theoremstyle{thmstyletwo}%

\theoremstyle{thmstylethree}%

\begin{document}

\title[Article Title]{Acoustically Driven Motion of Single Molecules at hBN Surfaces}


\author*[1]{\fnm{Hande N.} \sur{Açıkgöz}}\email{h.n.acikgoz@tudelft.nl}

\author[2]{\fnm{Kenji} \sur{Watanabe}}

\author[3]{\fnm{Takashi} \sur{Taniguchi}}

\author[1]{\fnm{Peter G.} \sur{Steeneken}}

\author*[1]{\fnm{Sabina} \sur{Caneva}}\email{s.caneva@tudelft.nl}

\affil[1]{\orgdiv{Department of Precision and
Microsystems Engineering}, \orgname{Delft University of Technology}, \orgaddress{\street{Mekelweg}, \city{Delft}, \postcode{2628CD}, \country{The Netherlands}}}

\affil[2]{\orgdiv{Research Center for Electronics and Optical Materials}, \orgname{National Institute for Materials Science}, \orgaddress{\street{1-1 Namiki}, \city{Tsukuba}, \postcode{ 305-0044}, \country{Japan}}}

\affil[3]{\orgdiv{Research Center for Materials Nanoarchitectonics}, \orgname{National Institute for Materials Science}, \orgaddress{\street{1-1 Namiki}, \city{Tsukuba}, \postcode{ 305-0044}, \country{Japan}}}

\abstract{Developments in single-molecule imaging and sensing techniques have enabled direct observation of biomolecular dynamics at interfaces, which has important implications for the fields of membrane biophysics, nanofluidic sensing and transport. However, integrating active surface transport, localization and control over molecule motion remains challenging, limiting further progress in the realization of single-molecule devices. Here, we introduce an integrated surface acoustic wave (SAW)-hexagonal boron nitride (hBN) platform as a contactless actuation platform for directional transport of biomolecules over hBN surfaces. hBN provides an atomically flat surface on which DNA can be confined and still retain in-plane mobility, whereas SAW actuation enables active and guided transport without requiring specific electrical, dielectric or magnetic properties of the biomolecules. Trajectory- and jump-based analyses show that increasing the SAW power leads to higher displacement bias in the streaming direction, resulting in transport velocities along increasingly more linearized trajectories. We thus demonstrate that SAWs can induce tunable directional transport of DNA molecules confined on hBN surfaces, providing a new approach for the on-chip manipulation of biomolecules in nanofluidic systems. Our findings have important implications for the development of advanced single-molecule sequencing platforms and other technologies for biomolecular manipulation and analysis. 
}


\maketitle

\section{Introduction}\label{sec1}

 A major objective in single-molecule techniques is to achieve deterministic manipulation of individual biomolecules. When confined to a surface, such a technique would enable molecules to be trapped, guided, and transported towards on-chip sensors, sorters and analyzers. Due to the molecular size scale, single-molecule dynamics are usually governed by random Brownian motion, making on-chip single-molecule measurements rely on random molecule-probe interactions, thereby lowering the yield and requiring longer measurement times for accurate readouts. 
Achieving control over spatial parameters, such as position and direction of motion, would make it possible to prolong interaction times, repeatedly interrogate the same molecule, perform sequential measurements with multiple probes, and ultimately transform random diffusion-driven detection into programmable sensing and manipulation.\\
\\
Conventional tweezing techniques for single particle manipulation are based on externally generated fields (e.g. magnetic \cite{Hodeib2016,Lipfert2014}, optical \cite{Bustamante2021}, electric \cite{Lee2004, Tian2024}, acoustic \cite{Collins2015}) applied to the medium in which the objects of interest are dispersed. The resulting force that an object within these fields experiences is dependent on its size, geometry, and relative properties with respect to the medium. While highly successful in the application of forces and torques to biomolecules, they have certain limitations when scaled down to the single-molecule regime. Magnetic tweezers require magnetic beads or target molecules with sufficient magnetic susceptibility \cite{Jin2026}, which most molecules intrinsically lack. They also rely on biomolecule attachment to a surface, thereby necessitating biochemical conjugation steps for tethering \cite{Lipfert2014}. 
Optical tweezers provide finer spatial control, yet the optical forces generated on nanometer-sized molecules become extremely small due to their limited scattering cross-section \cite{Bustamante2021}, and as a result require high laser intensities that may induce local heating or photodamage\cite{Landry2009}, or alternatively require attachment to micon-sized beads \cite{Newton2019}. Furthermore, optical tweezers are not suited for massively parallelized measurements since a single trapping beam typically manipulates only one molecule at a time, limiting scalability and throughput for large-scale single-molecule studies \cite{Whitley2017}.
Manipulation via electrophoretic and dielectrophoretic approaches rely on molecular charge and electric polarizability, respectively. Electrophoresis requires charged molecules\cite{Castro1995}, while dielectrophoresis produces significant forces only in the immediate vicinity of microelectrodes and are, as such, often used as a trapping mechanism \cite{Hlzel2005}. Additionally, both methods are affected by ionic screening and electrochemical effects in liquid environments. Therefore, these approaches limit large-area and long-range biomolecular transport. Hydrodynamic flow based techniques provide an alternative method to apply shear-induced drag forces on multiple surface-tethered double-stranded DNA \cite{yasunaga2026}, and have also been used to transport macromolecular PEG adsorbates along surfaces \cite{Velay2026}.  \\
 \\
Acoustic tweezers, which use sound pressure to indirectly apply forces to solids and liquids, have been highly effective in manipulating micron-sized objects \cite{Shen2024, UndvallAnand2021}. By using programmable harmonic acoustic fields, it is possible to generate reconfigurable particle traps with controllable spacing, as well as selective control over a particle's rotational motion \cite{yang2022harmonic}. Scaling this approach down to the molecular level, however, results in direct acoustic radiation forces becoming too weak to stably trap individual molecules against thermal fluctuations \cite{Kamsma2016}. Nevertheless, surface acoustic wave (SAW) have enabled reversible reorganization and patterning of DNA electrostatically bound to lipid bilayer membranes \cite{hennig2009dynamic, hennig2011dna}. On the other hand, acoustic streaming-based manipulation offers an alternative mechanism in which acoustically induced hydrodynamic flows transport particles indirectly through viscous drag forces \cite{Collins2017,Ma2017,skov2019three}. Compared to mechanically pumped hydrodynamic flows, SAW induced flows can be actuated electronically and locally on-chip, enabling faster flow modulation in a more compact device footprint. Moreover, the versatility in the number, shape and arrangement of the IDTs offers electronic reconfigurability of the flow properties and direction without modifying the fluidic circuitry \cite{li2025}. Despite lacking selectivity, SAW-driven approaches can simultaneously transport multiple objects largely independently of their electrical, magnetic, or optical properties, making them attractive for scalable and parallelized manipulation schemes, as well a compatible with a large range of materials and compounds. However, although SAW-driven acoustic manipulation approaches have been demonstrated for particle sizes down to $\sim$100\ nm \cite{Collins2017}, deterministic motion and positioning of single molecules over surfaces has not been realized using acoustic forces. \\
\\
Recent interest in hexagonal boron nitride (hBN) crystals and their interactions with biomolecules opened up new possibilities for on-surface single molecule analysis based on imaging, sensing and transport \cite{Zhang2025,orekhova2026}. hBN is a wide band gap, two-dimensional (2D) material that offers a chemically inert, non-fluorescent large-area surface free of dangling bonds. These features are particularly relevant for the analysis of surface-confined molecular systems, where physisorbed molecules remain accessible for optical interrogation while still undergoing stochastic Brownian diffusion \cite{Simon2024, Buskermolen2022}. Previous work showed that single-stranded DNA (ssDNA) and double-stranded DNA (dsDNA) molecules efficiently adsorb to hBN surfaces and undergo planar diffusion \cite{Shin2023,Slzle2024}. Although the diffusion behaviour appears random, single-molecule fluorescence localization and tracking revealed that ssDNA motion was confined by the hBN flake edges and line defects resulting in anomalous diffusion described by the mean squared displacement of $MSD=4D_A\tau^{\alpha}$. The anomalous diffusion coefficient $D_\alpha$ is found to be dependent on ssDNA length, ranging from approximately $0.02$ for 100-nt ssDNA to $0.16$ for 7-nt ssDNA. The exponent $\alpha$ also experiences a decrease with increasing DNA length, showing more pronounced subdiffusive behavior \cite{Shin2023}. hBN is thus a promising platform for nanoengineered transport and confinement of single molecules in an atomically flat landscape, while being fully compatible with standard fluorescence detection schemes. Yet, beyond analysis of in-plane motion, the potential for guiding single molecules towards on-chip sensing and sequencing elements, such as nanopores and fluorescence detectors, requires overcoming stochastic Brownian-like diffusion, which lacks directed motion control \cite{Cohen2008,Wang2014}. Specifically, we note the potential for a monolithic hBN  transport and sensing platform, which can harness, on one hand, atomically sharp hBN trenches as spatial guides, and on the other hand, nanoscale hBN optical emitters directly integrated in the surface as optical detectors \cite{shin2023single, Slzle2024}. Achieving guided, deterministic transport towards on-chip probes is, therefore, a key feature to unlock the potential of this nanofluidic system for molecular diagnostics.\\

\noindent
In this work, we introduce a hybrid surface acoustic wave (SAW)-hBN platform for the contactless directional transport and tracking of ssDNA. The device employs SAW actuation in fluid to control the direction of motion of ssDNA confined to an atomically flat hBN surface. Crucially, the hBN surface plays a dual role: it provides surface interactions sufficient to confine the biomolecules to the surface without requiring additional biochemical surface treatments, while lateral forces are sufficiently small to enable their in-plane motion. The surface confinement greatly simplifies studying the influence of forces on single-molecule dynamics because adsorption on hBN keeps the ssDNA focused on the imaging plane and the non-quenching property of hBN, which is unique among 2D materials, is directly compatible with high resolution fluorescence imaging techniques.
By tracking individual molecular trajectories with and without SAW actuation, we establish that acoustic effects modify the stochastic surface dynamics at the single-molecule level. The jump-displacement distributions reveal heterogeneous dynamics, consisting of a central population dominated by short displacements and a tail population containing larger jumps. The central population is comparatively more affected by localization uncertainty, whereas SAW-induced changes are more readily observable in the tail population. Under actuation, molecular motion develops an increasing directional bias along the streaming direction. With increasing SAW power, trajectories show a stronger net directional component superimposed on the underlying stochastic surface motion, resulting in progressively more persistent transport along the flow direction. Importantly, the generalized diffusion coefficient \(D_{\alpha}\) also increases with SAW input power, indicating that SAW actuation enhances molecular motion in addition to introducing a directional bias. These observations demonstrate that SAWs can bias the motion of surface-confined ssDNA toward directional molecular-scale transport while retaining some stochastic dynamic character, providing a pathway toward controlled manipulation of diffusing biomolecules on surfaces.

\section{Results and Discussion}\label{sec2}

Traveling wave SAW (TSAW) devices are fabricated on $128^\circ$ YX lithium niobate (LiNbO$_3$) substrates of 500 $\mu$m thickness via standard cleanroom techniques. The device is composed of one interdigital transducer (IDT) port that generates a 20 $\mu$m acoustic wavelength (Fig. 1a).  Details of the fabrication are given in the Methods. Subsequently, a multilayer hBN flake is mechanically exfoliated on PDMS and transferred to the SAW chip active area at around $100\ \mu$m distance from the IDT via a viscoelastic stamping method. An optical image of a stamped hBN flake in a representative SAW device is given in Fig. 1b. 
\\

\begin{figure}[H]
  \centering

 \includegraphics[width=0.99\textwidth]{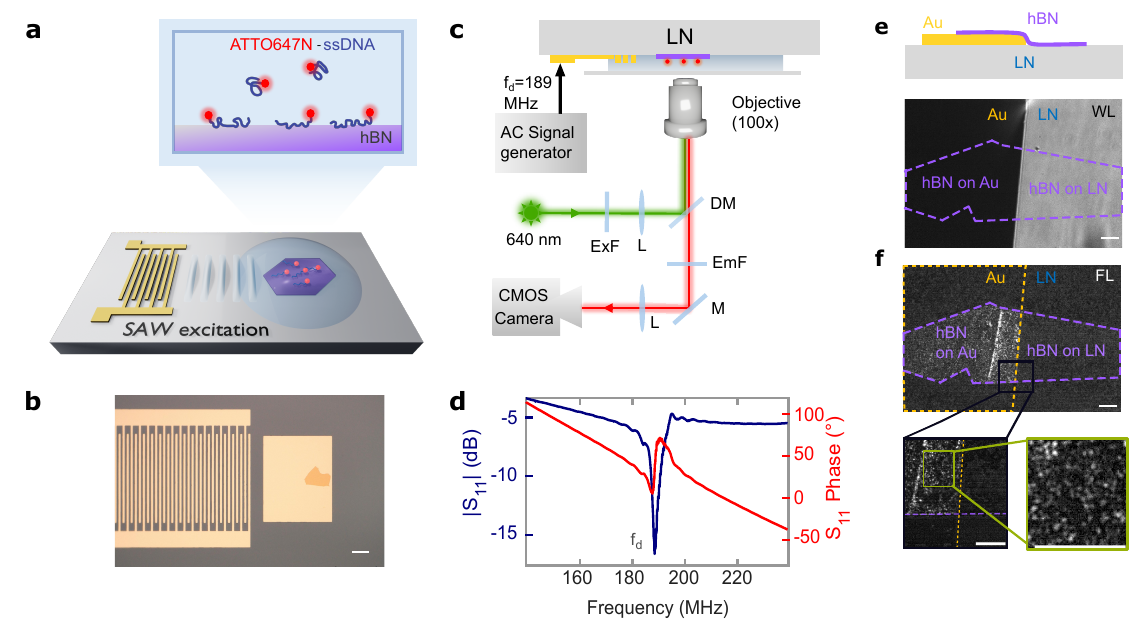}

  \caption{Experimental setup and single-molecule tracking method. a) Schematic of the SAW device with ATTO647N-labelled ssDNA molecules on an hBN flake. b) Optical image of a representative SAW device and gold pad with an hBN flake. The scale bar represents 40 $\mu$m. c) Inverted wide-field fluorescence microscopy setup for the detection of ssDNA molecules on the SAW chip during actuation with an AC signal.  d) $S_{11}$ reflection parameters of the SAW device. e) Schematic (top) and white light (WL, bottom) microscope image of an hBN flake layered partly on top of thin gold film and on bare LN surface. f) Fluorescence (FL) image of the same area with ATTO647N-ssDNA molecules on top. Scale bars in e,f represent 10 $\mu$m. }
  \label{fig:Experimental setup}
\end{figure}
\noindent
An inverted widefield microscope equipped with a 640 nm laser source, 100x magnification oil-immersion objective (NA 1.49) and a CMOS camera (Fig. 1c) is used to detect ATTO647N-ssDNA molecules. After introducing ssDNA onto the SAW chip, the device is placed on the microscope stage. The resonance frequency of the SAW devices is 189 MHz, as extracted from the $S_{11}$ parameter in a VNA measurement (Fig. 1d). Acoustic actuation is implemented by electrically connecting the SAW chip to an AC signal generator that supplies an AC signal at the 189 MHz drive frequency while imaging the fluorescence signal on the hBN surface. \\
\\
To enhance the intensity of the fluorescence signal from the labelled biomolecules, a thin gold pad was patterned and deposited along with IDTs on the wave propagation area and an hBN flake was positioned on top of it. A clear difference is observed between the detected fluorescence signal from ATTO647N-ssDNA on hBN/LN and on hBN/Au/LN, with the latter displaying higher intensities (Fig. 1e,f). FDTD simulations support this observation by showing fluorescence enhancement of a point source located at the hBN/water interface in the presence of an underlying Au layer (details in Supporting Information SI 1). \\
\\
Next, we simulated the properties of a simplified TSAW device and the resulting fields using a 2D COMSOL Multiphysics model. Fig. 2a shows the schematic of the model with 4 IDT fingers on a lithium niobate substrate and a water domain on top. The IDT fingers are actuated by an AC signal at 189 MHz driving frequency, resulting in a propagating surface wave. Out-of-plane vibrations at the solid-liquid interface lead to the leaking of the acoustic momentum into the liquid with Rayleigh angle $\theta_R$, which is defined by $\theta_R = \sin^{-1}(c_{\mathrm{w}}/c_{\mathrm{LN}})\approx 22 ^\circ$, where $c_w \approx$1481\ m/s and $c_{LN}\approx$3990\ m/s are the wave speeds at the water and LN domains, respectively.  A 1 V peak amplitude of alternating  voltage  is applied to the edge segments of $\lambda/4=5\ \mu$m size with the same $\lambda/4$ spacing representing the IDT fingerpairs. Details of the simulation model are given in SI 2. The electric potential distribution in the liquid domain (Fig. 2b) shows that the SAW-induced mechanical displacement is accompanied by a coupled electric potential field, with the potential gradient becoming less pronounced with increasing distance from the electrodes. Due to the applied alternating field at 189 MHz, the time-averaged electrophoretic motion is expected to be negligible. Dielectrophoretic effects are also expected to be limited away from the electrode edges  because of the weak local field gradient.\\
\\
\noindent
AC actuation of the IDT electrodes generates a surface wave propagating in the $+x$ direction away from the IDT.  The resulting  Rayleigh wave is characterized by a dominant out-of-plane displacement component as shown in the substrate cross-section in Fig.2 c. Part of the traveling wave energy leaks into the liquid domain with angle $\theta_R$ and produces the first order acoustic pressure field given in the top-half of Fig. 2c. The pressure field in the inset $I_{2}$ shows the close alignment of the surface displacement and resulting acoustic field nodes. Attenuation of this radiated acoustic field transfers momentum to the liquid and provides the body-force distribution that drives the second-order acoustic streaming pattern shown in Fig.2 d. The TSAW induces streaming fields in the wave propagation direction and the flow field is bounded by the no-slip boundary at the top, indicating the coverslip at  $80 \ \mu$m distance. The streaming velocity field near the LN–water interface is shown in the right panel of Fig. 2d. Close to the boundary, the velocity vectors are inclined relative to the surface, consistent with oblique acoustic leakage into the liquid, while farther from the interface the local flow becomes more Couette-like, with a predominantly tangential velocity profile that increases with distance from the surface. Although the hBN and gold were not included in this simulation we do not expect them to have a large effect on the acoustic flow profiles.\\

\begin{figure}[]
    \centering
    \includegraphics[width=0.8\linewidth]{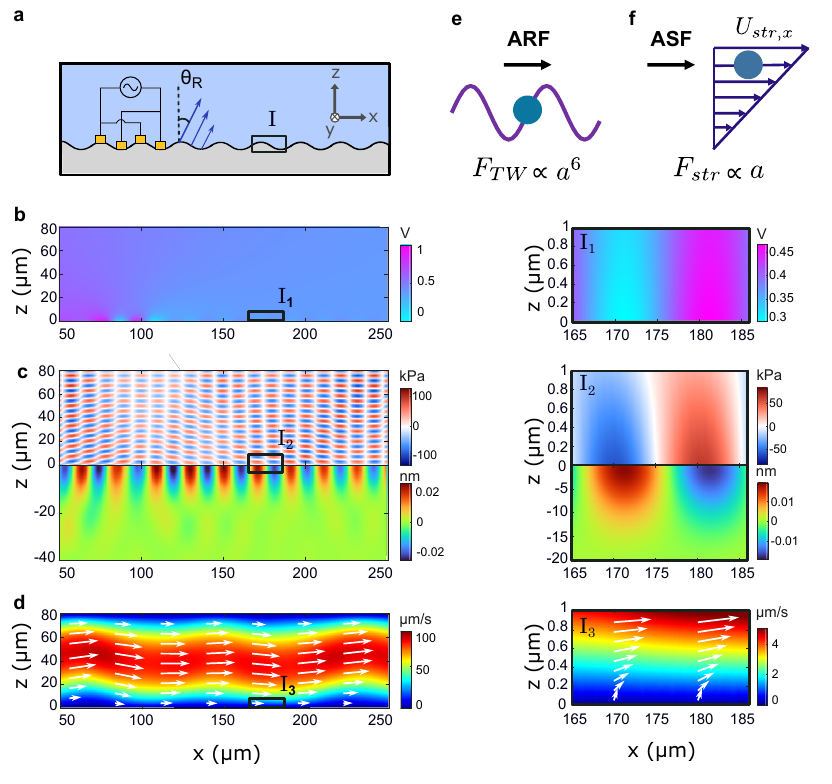}
    \caption{{Simulations of a simple TSAW device. a) Schematic of the simulation model representing a TSAW device with four IDT fingers on the LN substrate actuated with an AC signal. Regions of interest are indicated by boxes $I_{1-3}$. The corresponding magnified views are shown in panels (b-d) on the right. b) Resulting electric potential field in the liquid domain of $80 \ \mu$m of thickness in $z$. c) Vertical surface displacement amplitude due to the applied AC voltage on the IDT fingers and resulting first order acoustic pressure field in the liquid. d) Acoustic streaming flow velocity magnitude and direction. e) Schematic of a particle of radius $a$ in a travelling acoustic wavefield and induced acoustic radiation force (ARF) in the wave propagation direction. f) Schematic of the same particle located near a no-slip boundary in an acoustic streaming flow and acoustic streaming related drag force (ASF).}}
    \label{fig:Simulations of a simple TSAW device}
\end{figure}
\noindent
An object placed in a travelling acoustic wavefield will experience a time-averaged acoustic radiation force (Fig. 2e) due to the transfer of acoustic momentum. For a spherical particle with significantly smaller radius $a$ than the wavelength, $ka \ll 1$, the travelling-wave radiation force in the propagation direction is given as \cite{Hasegawa1969, Gokani2026}: 
\begin{equation}
\mathbf{F}_{\mathrm{TW}}=\frac{4\pi}{9}
E_{\mathrm{ac}}\,k^4a^6\,\Phi_{\mathrm{TW}}\,
\hat{\mathbf{k}},
\label{eq:TW_force}
\end{equation}
where $\hat{\mathbf{k}}$ is the unit vector in the direction of wave propagation, $E_{\mathrm{ac}}$ is the time-averaged acoustic energy density, $k=2\pi/\lambda$ is the acoustic wavenumber, and $\Phi_{\mathrm{TW}}$ is a dimensionless travelling-wave contrast factor determined by the density and compressibility mismatch between the particle and the surrounding fluid (SI 3). As the particle size becomes smaller, the acoustic radiation force significantly reduces due to the $a^6$ dependence, and becomes negligible compared to viscous drag or acoustic streaming-induced advection for nano-sized objects like ssDNA. When a similar particle is exposed to an acoustic-streaming flow, it experiences a viscous drag force, here referred to as the acoustic streaming force (ASF) (Fig. 2f), due to the relative motion between the particle and the surrounding fluid \cite{Stokes1851}.  
\begin{equation}
\mathbf{F}_{\mathrm{str}}
=
6\pi\eta a
\left(
\mathbf{U}_{\mathrm{str}}-\mathbf{U}_{p}
\right),
\label{eq:ASF}
\end{equation}
where $\eta$ is the dynamic viscosity of the fluid, $\mathbf{U}_{\mathrm{str}}$ is the local acoustic-streaming velocity, and $\mathbf{U}_{p}$ is the particle velocity.  Since the streaming induced drag scales with $\propto{a}$, conversely to a direct traveling wave effect scaling of $\propto{a^6}$ in the Rayleigh regime ($ka\ll1$), streaming induced transport dominates at these small scales. Using a pressure amplitude and streaming velocity of $P_0 = 50~\mathrm{kPa}$ and $U_{\mathrm{str,x}} = 1~\mu\mathrm{m\,s^{-1}}$ respectively, together with estimated DNA-like material properties of $\rho_p = 1700~\mathrm{kg\,m^{-3}}$\cite{Arrighi1970} and $\kappa_p = 2\times10^{-10}~\mathrm{Pa^{-1}}$\cite{Liu2021} , Eqs.~(1) and (2) predict markedly different force magnitudes for a $10~\mathrm{nm}$-radius spherical particle. The traveling-wave acoustic radiation force is of the order of $F_{\mathrm{TW}} \sim 10^{-25}~\mathrm{N}$, whereas the streaming-induced viscous drag force is of the order of $F_{\mathrm{ASF}} \sim 10^{-16}~\mathrm{N}$. Consequently, reducing the particle size strongly suppresses direct acoustic momentum transfer while leaving streaming-mediated transport comparably effective at the nanoscale.\\
\begin{figure}[]
    \centering
    \includegraphics[width=0.5\linewidth]{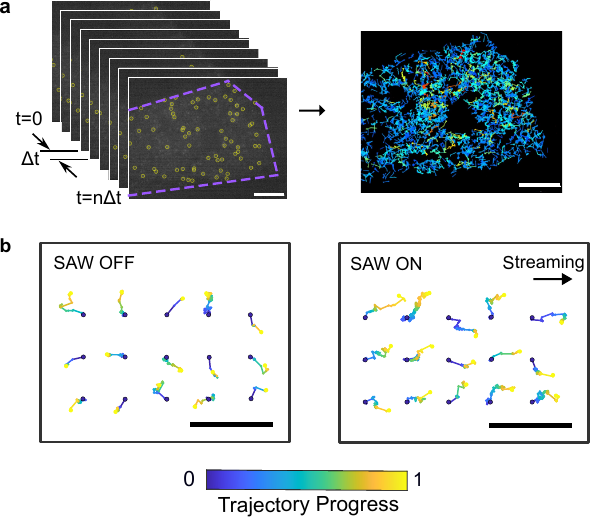}
    \caption{Single-molecule localization and tracking with SAW actuation. a) Detection of molecules at each time frame and resulting trajectories obtained by linking detected positions in the vicinity over time. b) Example trajectories from the experimental results with no SAW actuation (left) and with SAW actuation (right). The trajectories shown were selected among the relatively large displacement amplitude tracks in the respective datasets (OFF 1 and 20 dBm). In the SAW-actuated examples shown here, the trajectories are oriented along the streaming-flow direction as indicated by the yellow ends of the trajectory being located right of the blue starting locations. The scalebars represent 10 $\mu$m.}
    \label{fig:Single-molecule tracking}
\end{figure}
\noindent

\noindent
Time series measurements were recorded with 20 ms frame rate for 50 seconds with less than 9 mW laser power to limit bleaching of the fluorophores. Datasets were recorded for SAW OFF and SAW ON conditions, with the latter actuated at different input powers up to 23 dBm. Two OFF conditions were recorded, where OFF 1 refers to the initial recording before all the SAW ON conditions and OFF 2 refers to the last SAW OFF measurement after the power is turned off. The recorded time data images were then analysed with the ImageJ software \cite{Schindelin2012Fiji} and Trackmate plugin to localize the fluorophores and obtain trajectories (Fig. 3a). We filtered out trajectories with less than 5 detected positions to ensure that only trajectories containing sufficient positional information for reliable quantification of the nanomotion properties were included.\\
\\
\noindent
Seven single-molecule datasets were obtained as a function of SAW input power (OFF 1, 5, 10, 15, 20, 23 dBm, OFF 2). As the input power increases, more acoustic energy leaks into the liquid and will result in faster streaming flows. First, the detected trajectories were analyzed according to their trajectory displacements. For a trajectory $j$, the corresponding displacement, $\Delta R_j=(\Delta X_j,\Delta Y_j)$, is calculated based on the distance between the first and the last point of the trajectory (SI 4). In the absence of any external actuation or effect, random DNA motion leads to a distribution of  $\Delta R_j$ components $\Delta X_j$ and $\Delta Y_j$ centered around 0, while with increasing SAW input powers, and therefore higher streaming velocities, we observe an increase in the trajectory displacements in the flow direction. Example trajectories with and without SAW actuation are given in Fig. 3b, demonstrating how molecular motion aligns in the streaming flow direction.  \\

\begin{figure}[]
    \centering
    \includegraphics[width=1\linewidth]{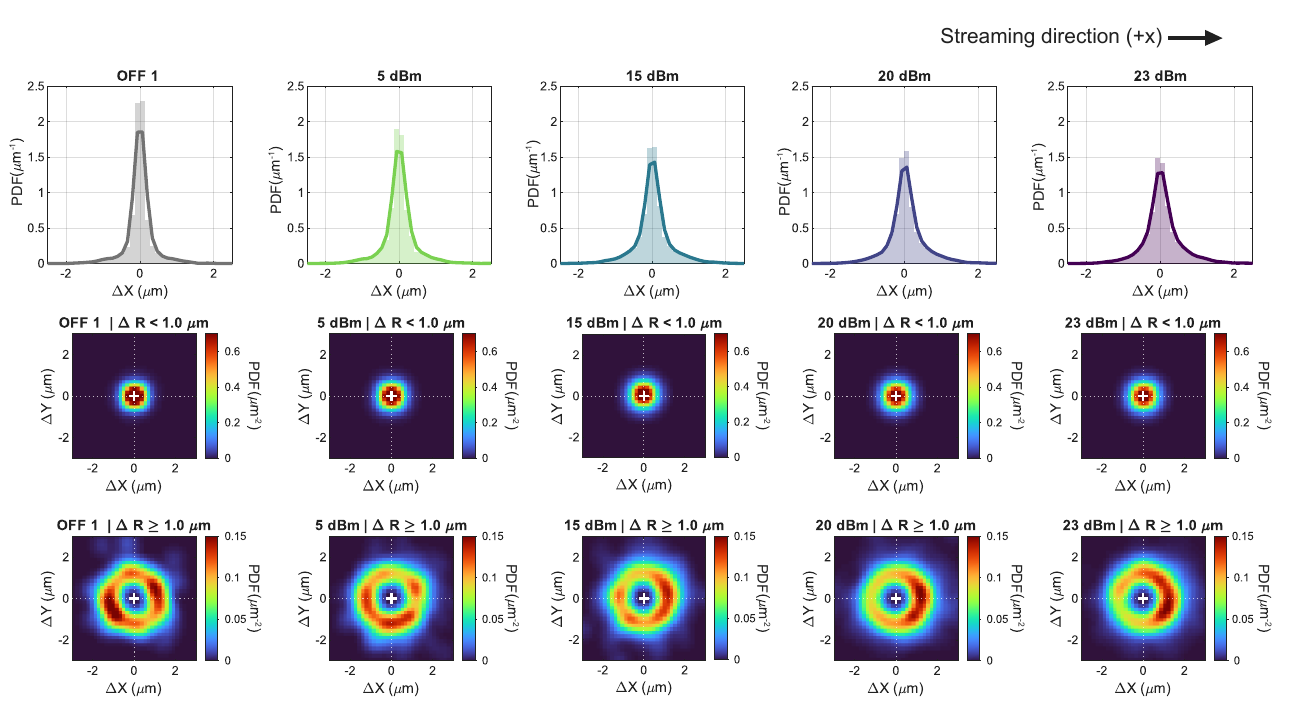}
    \caption{SAW-driven directional bias in nanomotion.  
a) Histograms of trajectory net \(x\)-displacements at different SAW input powers: OFF, 1, 5, 15, 20, and 23 dBm. The central peak remains near zero, while the distribution broadens with increasing SAW power; the directional bias is subtle in the full distribution since it is mainly carried by the tails.  b) Two-dimensional probability-density maps for trajectories with \(\Delta R<1~\mu\mathrm{m}\), showing a population concentrated near the origin.  c) Corresponding maps for \(\Delta R\geq1~\mu\mathrm{m}\), where the increasing asymmetry and extension toward \(+x\) clearly reveal the SAW-induced directional bias.}
    \label{fig:Trajectory level bias}
\end{figure}

\noindent
Fig. 4 a illustrates the histograms of trajectory displacements in the $x$-direction for 5 datasets, where two additional datasets are given in Fig. S3. The histograms remain centered close to $\Delta X=0$,  while the tail distributions become slightly asymmetric toward the streaming direction under SAW actuation. Consistently,  for trajectories with net displacements $\Delta R_j\geq1 \ \mu$m, the fraction of positive $x-$ displacement 
 $(\Delta X_j>0)$ increased from $47.1 \%$ under OFF 1 condition to $52.1 \%,\ 54.9 \%,\ 54.7 \%,\ 59.6 \%$ and $58.2 \%$ with increasing SAW input power, and reduces back to $48.7 \%$ for the OFF 2 condition . In parallel, the number of trajectories with $\Delta R_j\geq1 \ \mu$m increased substantially under SAW actuation, indicating an increased occurance of long-displacement trajectories. We also note that the probability density function (PDF) curves broaden and their central peak values decrease with SAW power, indicating an overall increased diffusion rate in all directions due to SAW actuation. In the OFF 2 conditions, the curve reverts to a narrower, sharp peak. The mean signed trajectory displacements along $x$, $\langle\Delta X\rangle $ increase from -0.0029 $\mu$m in the OFF 1 condition to $0.05\ \mu$m at 23 dBm with increasing SAW power.  Nevertheless, the overall distribution peak remained centered around zero and a modest directional bias is observed in trajectory-level metrics, indicating the stochastic motion remains dominant for most trajectories with a SAW-induced drift effect superimposed on the stochastic behavior. Fig.4 b,c show 2D probability density functions of trajectory displacements for trajectories  with net displacement amplitudes $\Delta R<1 \ \mu$m  and  $\Delta R\geq1 \ \mu$m, respectively. The 2D heatmaps show that the directional asymmetry becomes more apparent under SAW actuation, particularly for trajectories with larger net displacements. Figure 4 c is representative of the most mobile molecules and reveals a clear change in the displacement distribution. Under the OFF 1 condition, the trajectories are distributed more isotropically, resulting in a nearly circular displacement pattern. Upon SAW actuation, the distribution becomes increasingly biased toward the $+x$ direction, while the population extending toward $-x$ is reduced. This produces a progressively more asymmetric, lopsided distribution at higher SAW powers, consistent with an increasing directional bias along the acoustic-streaming direction.  Histogram and 2D heatmaps of remaining data (i.e. 10 dBm and OFF 2) is given in SI 5.\\
\begin{figure}[]
    \centering
    \includegraphics[width=1\linewidth]{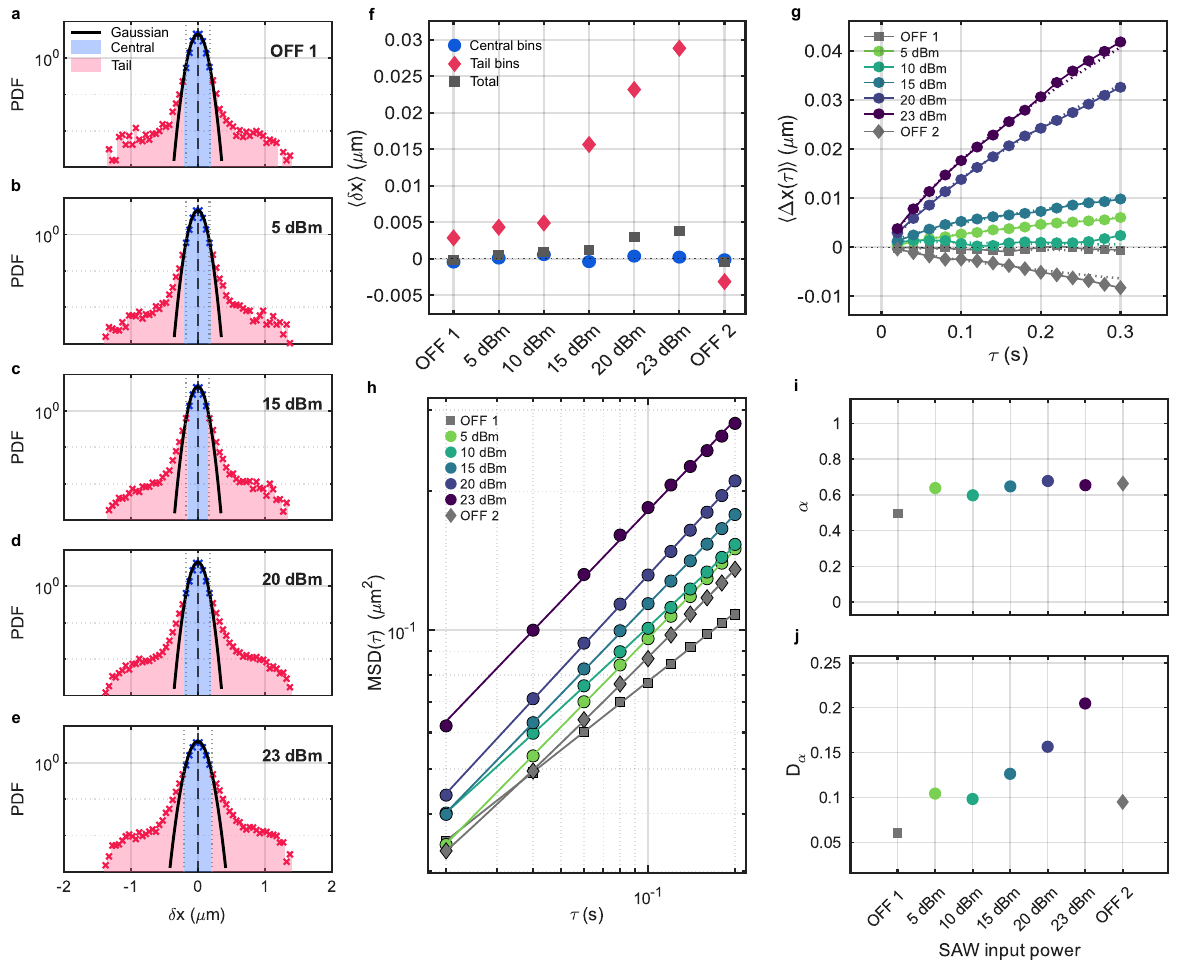}
    \caption{SAW effects on single DNA motion. a-e) Probability density function for signed frame-to-frame $x$-jumps, $\delta x$, at increasing SAW input powers. The central bins are identified from a Gaussian fit to the central part of the distribution; jumps outside these limits are classified as tail bins. f) Mean \(\delta x\) for the central-bin, tail-bin, and total jump populations as a function of SAW input power. g) Mean lag-dependent displacement $\langle \Delta x(\tau) \rangle$ for all datasets, dotted curves show the  fits used to estimate drift velocity. h) Mean squared displacement data as circles and fitted anomalous diffusion function $MSD=4D_\alpha\tau^\alpha$ as continuous line in dataset colors. i) Fitted $\alpha$ and fitted $D_\alpha$ values as a function of SAW input power.}  
    \label{fig:SAW effects on single DNA motion}
\end{figure}
\\
\noindent
While trajectory-level metrics quantify the overall transport response of the molecular population, they overlook the heterogeneity in a single trajectory which may be comprised of different dynamic states during the trajectory duration \cite{Simon2024}. We therefore further analysed individual jumps between two consecutive positions, 20 ms apart, in each trajectory and obtained jump displacement, $\delta x$, distributions, including frames with $\delta x=0$. Distributions for all datasets at different SAW input powers (Fig.5 a-e) reveal the existence of at least two states: a  central state characterized by short displacements around zero (blue), and a mobile tail state comprised of longer displacements (pink). Central and tail populations were defined based on a fitted Gaussian curve in the center with mean and standard deviation of $\mu_c$ and $\sigma_c$ respectively. Limits defining central bins are defined as [$\mu_c-2\sigma_c,\ \mu_c+2\sigma_c $], and the population outside of this range is defined as the belonging to the tail. We note that the central population also has a comparatively large contribution of localization noise as the experimentally estimated frame-to-frame displacement uncertainty of $\sigma_{\delta x}\approx$80\ nm becomes comparable to the $\delta x$ values in the central limits, which range approximately $\pm 0.17\ \mu$m to $\pm 0.21\ \mu$m across the datasets. Therefore, SAW-induced changes would be more readily observable in the tail population, where larger jumps are less affected by localization uncertainty. \\
\\
Next, we determine the confinement ratios along the $x$-axis (CR$_x$), defined for each trajectory as $|\Delta X|$, divided by the accumulated absolute displacement along $x$. A highly linear trajectory will have a confinement ratio close to 1, while a random trajectory with frequent reversals yields a value closer to zero, and this parameter is therefore an indicator of persistence of trajectories along the $x$-axis. Comparing the datasets across the SAW power series, we find an overall increase of the mean $x$-confinement ratio $\langle CR_x \rangle$ under SAW actuation, rising from 0.239 for SAW OFF 1 to 0.244, 0.261, 0.268 and 0.285 at 5, 15, 20 and 23 dBm, respectively. In addition to the increase in linearity of the trajectories, the frame-to-frame $x$-jump ($\delta x$) displacement analysis provides further evidence of a directional bias. As shown in Fig. 5f, increasing the SAW power shifts the $\delta x$ distribution of the mobile population toward the flow direction $+x$, while the immobile population exhibits little to no change. Thus,  the mean $\langle\delta x\rangle$ also shifts toward $+x$, consistent with increasingly directional transport under SAW actuation.\\
\\
\noindent
To look further into the induced directional movement, the molecular motion can be considered as a superposition of a stochastic component $\Delta x^{st}(t)$ and a streaming-directed component $\Delta x^{d}(t)$ (SI4). Therefore, the $x$-displacement can be written as $\Delta x(t)=\Delta x^{st}(t)+\Delta x^{d}(t)$. 
For the subdiffusive motion observed here, the directed response is described using the same anomalous lag-time dependence as obtained from the MSD analysis \cite{bouchaud1990anomalous, chen2019subdiffusion, barkai1998generalized}, while the stochastic component approaches zero when ensemble-averaged, $\langle\Delta x^{st}(t)\rangle\approx0$. Accordingly, the mean displacement can be described as $\langle\Delta x(t)\rangle\propto t^\alpha$ with $\alpha<1$.\\
\\
\noindent
 To quantify this net directional motion, the lag-time dependence of the mean $x$-displacement, $\langle\Delta x(\tau)\rangle$, was analyzed. Here, $\Delta x(\tau)$ denotes the displacement accumulated over a lag time $\tau=m\Delta t$, where $m$ is a positive integer and $\Delta t=20~\mathrm{ms}$ is the experimental frame time. This lag-time displacement is distinct from the frame-to-frame jump $\delta x$ used in Fig.5 f. At the first lag time, $\tau=\Delta t=20~\mathrm{ms},$ the two quantities are equivalent. Accordingly, the first data point in panel g corresponds to the ``Total'' population shown in panel f. Examining several lag times enables identification of persistent directional motion, which appears as a systematic change in $\langle\Delta x(\tau)\rangle$ with increasing $\tau$. For subdiffusive transport, the mean displacement is described as $\langle\Delta x(\tau)\rangle=A_x\tau^\alpha+b_x$, where $\alpha$ is fixed to the value obtained independently from the drift-corrected MSD analysis for the corresponding SAW condition, so that the lag dependence of the directional response is constrained by the independently measured subdiffusive exponent. The effective lag-dependent drift velocity in the $x$ direction is then obtained as $v_{\mathrm{eff},x}(\tau)=\alpha A_x\tau^{\alpha-1}$ \cite{barkai1998generalized}. Here, $b_x$ accounts for a possible small displacement offset and does not contribute to the lag-dependent velocity. Under the generalized Einstein linear-response assumption, the fitted amplitude $A_x$ would be related to the effective force as $A_x=D_\alpha F_x/(k_{\mathrm B}T)$, and therefore depend on both the anomalous diffusion coefficient and the applied effective force. The generalized Einstein relation is used here separately for the effective-force estimate, under the weak-bias linear-response assumption discussed in SI7. Fig.5 g shows the ensemble-averaged lag-time displacements for the different SAW input powers with fitted functions. Fits are performed with the first 10 data points, and the resulting $x$- and $y$-directed velocities $(v_{\mathrm{eff},x}(\tau))$ and $(v_{\mathrm{eff},y}(\tau))$ are given in Table 4.1 below.

\begin{table}[ht]
\centering
\caption{Lag-time-dependent effective drift velocities along the $x$ and $y$ directions.}
\label{tab:veff_xy}
\begin{tabular}{lccc}
\hline
Condition & $\alpha$ &
$v_{\mathrm{eff},x}(\tau)=\alpha A_x\tau^{\alpha-1}$ &
$v_{\mathrm{eff},y}(\tau)=\alpha A_y\tau^{\alpha-1}$ \\
\hline
OFF 1  & 0.49 & $-0.0007\,\tau^{-0.51}$ & $-0.0091\,\tau^{-0.51}$ \\
5 dBm  & 0.64 & $\phantom{-}0.0094\,\tau^{-0.36}$ & $-0.0069\,\tau^{-0.36}$ \\
10 dBm & 0.60 & $-0.0010\,\tau^{-0.40}$ & $-0.0017\,\tau^{-0.40}$ \\
15 dBm & 0.65 & $\phantom{-}0.0141\,\tau^{-0.35}$ & $-0.0031\,\tau^{-0.35}$ \\
20 dBm & 0.68 & $\phantom{-}0.0558\,\tau^{-0.32}$ & $-0.0040\,\tau^{-0.32}$ \\
23 dBm & 0.65 & $\phantom{-}0.0645\,\tau^{-0.35}$ & $\phantom{-}0.0227\,\tau^{-0.35}$ \\
OFF 2  & 0.66 & $-0.0106\,\tau^{-0.34}$ & $-0.00003\,\tau^{-0.34}$ \\
\hline
\end{tabular}
\end{table}

 \noindent
 Although the response is not strictly monotonic at intermediate powers, the results indicate an overall increase in the directed velocity component in $x$ with increasing SAW input power, with the largest positive constant $\alpha A_x$ at 23 dBm.  This behavior is also consistent with the reported literature on flow-driven transport at solid-liquid interfaces  \cite{Shankla2019, Velay2026}, where the transport velocity at an interface can be tuned with flow velocity. Additionally, the initial OFF 1 the final OFF 2 measurement show a small negative drift components in $y$ and $x$  which may reflect a weak background drift of the sample or imaging system. $v_{eff,x}$  does not show a strong response for the lower input powers and even gives a small negative constant at 10 dBm, whereas the increase from 15 to 23 dBm becomes progressively more pronounced. In comparison,  $v_{eff,y}$ remains relatively small compared to $v_{eff,x}$ for most power conditions showing that the dominant response is in the forcing direction. The increase in $v_{eff,y}$  at 23 dBm may reflect  power-dependent changes in local streaming pattern or spatial non-uniformity of the acoustic field.    \\ 
\\
\noindent
The MSD analysis further shows that the behavior is well described by an anomalous diffusion model, $MSD=4D_\alpha \tau^\alpha$. Since actuation also introduces a small directional drift, we additionally recalculated the MSDs after correcting for this drift contribution (SI 4) (Fig. 5h). Fig. 5j shows that the generalized drift-corrected diffusion coefficient $D_\alpha$ increases from 0.0605 at initial OFF 1 condition to 0.1043, 0.0982, 0.1263, 0.1565 and 0.2047 for 5, 10, 15, 20 and 23 dBm input powers, respectively. At the last OFF 2 condition, after the SAW excitation is removed, it was observed that $D_\alpha$ decreases back to 0.0950, confirming that the majority of the increase results from the SAW actuation. In contrast, in Fig. 5i, the anomalous diffusion exponent $\alpha$ remains within a relatively narrow range of 0.49-0.66, indicating a persistance of the sub-diffusive state. The obtained $D_\alpha$ and $\alpha$ values, especially in the absence of the SAW-induced effects, are comparable in magnitude to reported values for similar length ($\sim 35$ nt) ssDNA motion on hBN surfaces \cite{Shin2023}. Differences in values may also arise from the differences in the buffer composition and ionic strength, which can modify electrostatic screening, ssDNA conformation and molecule-surface binding \cite{ibrahim2023}. \\
\\
\noindent Within this experiment, the $MSD=4D_\alpha \tau^\alpha$ increases with increasing SAW power. This increase is reflected in a higher anomalous diffusion coefficient \(D_\alpha\), while the anomalous exponent \(\alpha\) remains approximately unchanged, indicating that SAW actuation enhances the extent of molecular displacement while preserving the subdiffusive character of the motion. In addition, SAW introduces a directional bias along the propagation direction. The observed enhancement in DNA motion with SAW can be explained by the SAW-induced modulation of the diffusion energy landscape. It has been shown theoretically and experimentally that SAWs can enhance surface diffusion of atomic and molecular adsorbants on a surface through transient strain-induced modification of diffusion barriers and local temperature variations \cite{wu2013acoustic, shugaev2015}. The theoretical study predicts a quadratic scaling of the diffusion amplitude with SAW-induced strain amplitude, in the small amplitude regime \cite{wu2013acoustic}. Considering that the exponent $\alpha$ values within this experiment remain at around $0.6$ across datasets, we compared the relative change in $D_\alpha$ across SAW powers. After the drift velocity correction and OFF-state baseline correction, the enhancement factor $D_\alpha/D_{\alpha,base}-1$ shows an approximately quadratic dependence with applied peak voltage, consistent with the proposed strain-induced diffusion enhancement model \cite{wu2013acoustic}. \\
\\
\noindent
Since the system can be considered under weak forcing, the generalized Einstein relation can be applied as an effective linear-response description \cite{bouchaud1990anomalous}. The generalized Einstein relation is used here as an effective model to estimate the force acting on the subdiffusive DNA motion. Its use is supported by the relatively small directional bias compared with the stochastic displacement fluctuations, with $|\langle\Delta x(\tau)\rangle|/\sqrt{\langle\Delta x^2(\tau)\rangle_0}\sim0.1$ at $\tau=0.4$ s. The force is calculated using the corresponding drift-corrected MSD measured at each SAW input power, so that SAW-induced changes in the stochastic motion are taken into account. In addition, the inferred force becomes approximately independent of lag time after a short transient time, supporting the use of the relation as an effective force estimate. However, deviations from the generalized linear-response description may occur outside the weak-bias regime or when surface heterogeneity and interaction dynamics produce nonlinear response \cite{bouchaud1990anomalous}. By directly using the experimental mean displacement and corresponding drift-corrected 1D mean-squared displacement data along $x$ for each SAW input power, the temperature-normalized force along $x$, $F_x/(k_{\mathrm B}T)$, can be estimated (see SI7). As expected, the inferred normalized force overall increases with the increased bias in $\langle \Delta x(\tau)\rangle$. The larger values are observed at the higher SAW input powers, with a maximum of approximately $0.45~\mu\mathrm{m}^{-1}$ obtained at 20 dBm. 
Assuming room temperature, $T=298~\mathrm{K}$, this value corresponds to an effective force of approximately $1.8\ \mathrm{fN}$.\\
\\
\noindent This force value, inferred from the experimental data under the generalized Einstein relation, is additionally compared with an independent estimate of the hydrodynamic drag on a particle at a wall subjected to Couette flow. Using the simulation result from a full-device model with $N_p=40$ finger pairs, the drag force induced by streaming at 23 dBm can be estimated as $F_{\mathrm{st}}\sim0.46\,\mathrm{fN}$, while a possible dielectrophoretic contribution is estimated to be $F_{\mathrm{DEP}}\sim0.002\,\mathrm{fN}$, approximately two orders of magnitude lower than the streaming contribution. This shows that the experimentally inferred force scale assuming room temperature is of a comparable order of magnitude to the independently estimated hydrodynamic drag generated by the SAW actuation, supporting acoustic streaming as the likely main contributor to the observed directional drift, while possible nonlinear electrical contributions cannot be excluded.\\
  \\
\section{Conclusion}\label{sec3}

In this work, we introduced a hybrid surface acoustic wave (SAW)-hexagonal boron nitride (hBN) platform for the contactless manipulation of ssDNA molecules  confined to atomically flat hBN surfaces. By using single-molecule fluorescence tracking, we investigated the effect of acoustic actuation on ssDNA motion through trajectory-based metrics and individual jump statistics. Our results demonstrate the existence of heterogeneous dynamics in both trajectories and individual jumps, with a coexistence of mobile and immobile DNA populations in which the former show a stronger response to SAW actuation. In addition to an increased overall mobility, both net trajectory displacements and individual jumps show increased bias towards the acoustic wave propagation direction, revealing enhanced directionality with increasing SAW power. These biases in direction and velocity suggest that SAW-induced forces progressively align molecular transport in the traveling SAW wave direction, resulting in higher drift velocities and more linearized trajectories. \\
\\
\noindent
The device presented here provides a fully integrated actuation and imaging approach that leverages SAW-induced flows to control the transport direction and velocity of biomolecules on-chip. By integrating SAW devices into single-molecule imaging and sensing platforms, this acoustofluidic effect can be exploited to deterministically transport analytes towards optical or electronic sensors. We expect additional functionality of this platform from the use of the hBN substrate. Specifically, previous reports on molecular dynamics simulations on hBN/graphene heterostructures established that biomolecule-hBN interactions facilitate the denaturation and linearization of DNA and proteins, which are essential for base-by-base or sequential amino acid recognition, without requiring additional biochemical steps such as SDS treatments \cite{Luan2019}. Thus, we envisage that nanopores and optical emitters fabricated directly in hBN could provide sequence-sensitive readout via either ionic current or fluorescence intensity modulations \cite{Wells2012, Wilson2016}. We also note the compatibility of our device and material system with interferometric scattering microscopy (iSCAT) \cite{Slzle2024}, a label-free technique, where longer tracks could be monitored, thereby extending observation and analysis time window. Ultimately, combining the surface properties of hBN with tunable, local acoustic forces, can enable platforms where molecules are stretched and guided towards nanoscale probes, contributing to unlocking new approaches for next-generation on-chip biomolecular analysis. 

\bmhead{Supplementary information}
Supplementary information is provided for Fluorescence enhancement simulations, COMSOL simulations, Acoustic radiation force due to travelling wave, Trajectory analysis, Remaining datasets and $y$-direction analysis, SAW-enhanced diffusion and Generalized Einstein relation and force estimations

\bmhead{Acknowledgements} H.N.A. and S.C. acknowledge funding from the European Union’s Horizon 2020 research and innovation program (ERC StG, SIMPHONICS, Project No. 101041486). K.W. and T.T. acknowledge support from the CREST (JPMJCR24A5), JST and World Premier International Research Center Initiative (WPI), MEXT, Japan.

\section{Methods}\label{sec4}
\textbf{SAW device fabrication.}
  TSAW IDT design is comprised of $N_p=40$ fingerpairs with $\lambda/4=5\ \mu$m finger width and spacing. Finger overlap is $220\ \mu$m. $128^\circ$ Y-cut X-propagating Lithium Niobate (LN) wafers of 500 $\mu m$ thickness were diced into 10 mm x 5 mm pieces. The LN chips were first spin-coated with HMDS and baked at $200^\circ C$ for 2 min, then spin-coated with AZ ECI 3007 resist and baked at $90^\circ C$ for 15 min. The IDT and Au pad pattern were generated on the surface with UV lithography followed by the electron beam (e-beam) deposition of 5 nm Ti and 70 nm Au. The process was finalized with a liftoff step at acetone. The fabricated SAW devices were connected to a PCB by wirebonding. hBN flakes were exfoliated on PDMS from an exfoliation tape and the selected flake with a thickness of $\sim 35$ nm is picked up by a PDMS viscoelastic stamp and placed in the SAW active area, approximately $108\ \mu$m. \\ away from the IDT.\\
\textbf{Fluorescence imaging.}
A fluorescence microscopy setup (Nikon Eclipse Ti2) is used to record time series data from single molecules excited with $\sim$15\ mW laser power at 640 nm wavelength. The exposure time was set to 20 ms and the total recording time was 1 min. First, a droplet of buffer solution (10 mM TrisHCl, 100 mM KCl) was placed on the SAW chip and covered with a 130 $\mu$m thick coverslip. Then the SAW chip was placed on the microscope and the background fluorescence signal from the $\mathrm{LiNbO_3}$ substrate and hBN flake was extracted, serving as the baseline. The chip was mildly washed with MilliQ water before placing the droplet of DNA solution in buffer containing 200 pM ATTO647N-labeled 29-nt ssDNA of sequence ATTO647N-CGGCAAATCCCTTTAGGGGGTGGTTCGA, 10 mM TrisHCl and 100 mM KCl.  After 10 min incubation time required for the ssDNA molecules to adsorb on the hBN surface, the SAW chip was again mildly washed with MilliQ water to remove excess ssDNA molecules suspended in solution. Lastly, a droplet of buffer solution was placed on top of the device again, and covered with a coverslip.\\
\textbf{ Single-molecule tracking.} The ImageJ \cite{Schindelin2012Fiji} Trackmate plugin was used to detect the fluorophore signals and track their positions over time. A difference of Gaussian (DoG) detector was used to detect molecules with the particle-detection diameter set to 0.9 $\mu$m, and the detection threshold was set to 0.2. Median filtering was enabled prior to detection to reduce image noise, and sub-pixel localization was enabled. Detected spots were subsequently used for single-molecule trajectory analysis. The detected spots were linked into trajectories using the Simple LAP tracker, with a maximum linking distance up to 1.4 $\mu$m to capture long distance jumps. Temporary detection gaps were bridged using a maximum gap-closing distance of 1.0 $\mu$m over a maximum of 2 consecutive frames.  The localization error is estimated from the tracking of a stationary emitter at the edge of the flake, under the same imaging and tracking conditions. The standard deviation of the measured position is taken as the single-frame localization precision $\sigma=(\sigma_x,\sigma_y)$ while frame-to-frame displacement uncertainty is estimated as $\sqrt{2}\sigma=(\sqrt{2}\sigma_x,\sqrt{2}\sigma_y)$=(78.28\ nm, 68.49\ nm). 

\bibliography{Ref}
\end{document}